# Investigating society's educational debts due to racism and sexism in student attitudes about physics using quantitative critical race theory

Jayson M. Nissen,[1] Ian Her Many Horses,[2] and Ben Van Dusen[3]
[1]Nissen Education Research and Design, Corvallis, Oregon 97333, USA
[2]School of Education, University of Colorado Boulder, Boulder, Colorado 80309, USA
[3]Department of Science Education, California State University Chico, Chico, California 95929, USA

The American Physical Society calls on its members to improve the diversity of physics by supporting an inclusive culture that encourages women and Black, Indigenous, and people of color to become physicists. In the current educational system, it is unlikely for a student to become a physicist if they do not share the same attitudes about what it means to learn and do physics as those held by most professional physicists. Evidence shows college physics courses and degree programs do not support students in developing these attitudes. Rather physics education filters out students who do not enter college physics courses with these attitudes. To better understand the role of attitudes in the lack of diversity in physics, we investigated the intersecting relationships between racism and sexism in inequities in student attitudes about learning and doing physics using a critical quantitative framework. The analyses used hierarchical linear models to examine students' attitudes as measured by the Colorado Learning Attitudes about Science Survey. The data came from the Learning About STEM Student Outcomes database and included 2170 students in 46 calculus-based mechanics courses and 2503 students in 49 algebra-based mechanics courses taught at 18 institutions. Like prior studies, we found that attitudes either did not change or slightly decreased for most groups. Results identified large differences across intersecting race and gender groups representing educational debts society owes these students. White students, particularly White men in calculus-based courses, tended to have more expertlike attitudes than any other group of students. Instruction that addresses society's educational debts can help move physics toward an inclusive culture supportive of diverse students and professionals.

DOI: 10.1103/PhysRevPhysEducRes.17.010116

## I. INTRODUCTION

The American Physical Society calls on its members to improve the diversity of physics by supporting an inclusive culture that encourages women and Black, Indigenous, and people of color (BIPOC) [1] to become physicists [2]. The physics community pursues this goal through several avenues including research on the barriers imposed on women and BIPOC students in their physics education. This research includes gender differences in grades [3–7], conceptual knowledge [8], and affective characteristics such as self-efficacy [4,9,10] and attitudes [4,6,11–14]. Research has also investigated sexual harassment and discrimination in physics courses [15,16]. Similar but less extensive work [17] has investigated these issues for BIPOC students in physics: grades [7], conceptual knowledge [6,18–21], classroom experiences [22–29], and affective characteristics [25]. In this study, we investigated the role of attitudes about learning and doing physics in contributing to the lack of diversity in physics.

Most professional physicists subscribe to a set of attitudes about what it means to learn and do physics [11,30,31]. In the current educational system, it is unlikely for a student to become a physicist if their attitudes about what it means to learn and do physics do not match those of most professional physicists [30,31]. Yet, physics educators tend to rank attitudes toward and appreciation of physics as less important student outcomes than conceptual understanding and problem-solving ability [32]. And, most college physics courses [33] and degree programs do not support students in developing these attitudes [30,31]. These two characteristics of physics education mean that physics degree programs filter out students who do not already hold these attitudes when they enter college physics courses [30,31].

Few studies have investigated demographic differences in attitudes about learning and doing physics. While several studies have shown more favorable attitudes and shifts in attitudes for men than for women [6,11–13] we only know of one study that discussed differences across race or ethnicity [14]. Traxler and Brewe [14] found gender and ethnic differences favoring men and overrepresented ethnicities (i.e., Asian and White [34]). They also found Modeling Instruction, an evidence based pedagogy with a focus on developing student attitudes, supported women and BIPOC students in developing more expertlike attitudes.

The disparities in outcomes for women and BIPOC students in physics courses result from systemic barriers in physics education. These barriers perpetuate the educational debts society owes these students [35]. Society has accrued educational debts that it owes to minoritized students through historical, sociopolitical, economic, and moral forms of inequalities [36]. In this investigation we examined an avenue by which the racist and sexist power structures within university physics courses perpetuate and increase the educational debts society owes women and BIPOC students through the denial of opportunities and resources to develop as physicists [10,37]. To better understand the role of attitudes in the lack of diversity in physics, we used a critical quantitative framework (QuantCrit) [38] to investigate the intersecting relationships between racism and sexism in inequities in student attitudes about learning and doing physics. We modeled society's educational debts due to racism, sexism, and their intersection in a multi-institutional dataset (18 institutions and 95 courses) collected using the Colorado Learning Attitudes about Science Survey (CLASS) [11] using hierarchical models. Our QuantCrit framework guided our work in an attempt to be antiracist and antisexist and counter racist and sexist uses of quantitative research in the past and present.

## II. RESEARCH QUESTIONS

To better understand the role of attitudes in the lack of diversity in physics we asked the following questions.
  1. To what extent are the effects of racism and sexism present in students' attitudes before and after taking algebra-based or calculus-based introductory mechanics courses?
  2. To what extent do differences in attitudes about physics perpetuate inequities in representation in physics?

Understanding the role of attitudes about learning and doing physics in the lack of diversity in physics can focus attention on this issue. This knowledge can motivate instructors and departments to adopt pedagogies and policies that support students from marginalized groups in developing expertlike attitudes about learning and doing physics.

## III. LITERATURE REVIEW

### A. Attitudes in physics

Physicists hold attitudes about the nature of science distinct from other science disciplines [11]. These cultural attitudes include valuing competition, individualism, and solitary practice, which are more costly for women and BIPOC students to adopt [29,39]. These cultural attitudes also include viewing physics as applicable to daily life, perceptions of personal effort supporting learning physics, and approaches to solving and understanding physics problems [40]. In this research, we focused on these latter beneficial attitudes because they support learning physics [41,42] and holding nearly all of these attitudes is a characteristic of professional physicists [30,31]. While most students in calculus-based mechanics courses hold a majority of these attitudes [33], few students hold enough of these attitudes to become professional physicists in the current system [31]. Physics courses and programs seldom support students in developing these attitudes; rather, physics courses tend to harm students' attitudes [12,13,33] and programs produce majors from the pool of students who already have the same set of attitudes [30,31]. While this filtering approach appears common, several physics curricula support students in developing expertlike attitudes [14,33].

While several studies have found gender differences in attitudes toward learning and doing physics [4,6,11–14], we are only aware of one study that also looked at differences across ethnicities [14]. Most, but not all, of these studies show differences favoring male students and White students. Studies investigating gender differences across a large collection of attitudes find gender differences favoring male students before instruction [4,6,11,13,14]. In most cases, the investigated physics courses maintained or increased the gender differences as the average attitudes either stayed the same or decreased for male and female students. However, these results are not universal. Traxler and Brewe [14] investigated differences between male and female students and students from overrepresented (i.e., White and Asian) and underrepresented ethnicities attitudes using the CLASS in introductory physics courses using Modeling Instruction. Modeling Instruction emphasizes attitudinal development and tends to improve students' attitudes about learning and doing physics [43], though this result has not been replicated at other institutions. They found slightly more expertlike attitudes for male students and for White and Asian students at the beginning of the term. Over the semester, attitudes increased for all groups of students. These increases maintained the small difference across ethnicities. However, the increase was much larger for female students than male students and at the end of the course female students had more expertlike attitudes on average than their male peers.

Good *et al.* [12] investigated attitudes about solving physics problems for male and female students in a traditional lecture-based physics course and a flipped course using their own instrument. They found female students in both courses started with more expertlike attitudes and maintained those more expertlike attitudes. The attitudes for their male peers decreased from before to after instruction in both courses. Good and colleagues' study differs from other referenced work because it focused explicitly on attitudes toward problem solving.

### B. Sexism in physics

Significant research efforts have focused on understanding the low representation of women in physics, which has remained at approximately 20% for the last 40 years [44]. Cheryan *et al.* [10] reviewed the literature on gender differences across the science, technology, engineering, and mathematics (STEM) domains and found that masculine cultures, gender differences in self-efficacy, and a lack of early educational experiences in the disciplines explained the lower rates of participation for women in physics, computer science, and engineering compared to biology, chemistry, and mathematics. Work in physics education research on gender differences in physics parallels Cheryan and colleagues findings. Madsen *et al.* [8] reviewed 26 studies on gender differences for conceptual learning in introductory physics courses. In first semester physics courses, Madsen and colleagues found that in the 26 studies on conceptual learning they reviewed the male students average pretest scores were always higher than female students average pretest scores (13% weighted average difference) and in most cases male students average post-test scores were also higher (12% weighted average). Most of the 26 studies did not find statistically significant differences in the learning across genders. Madsen and colleagues concluded the studies did not identify a single factor or solution for the gender differences, but that the gender differences are likely due to a combination of factors. This difference at the beginning of the term aligns with Cheryan and colleagues [10] findings about girls having fewer early educational opportunities in physics. Studies have also reported on gender differences in final course grades in introductory physics courses with mixed results. In some studies male students received higher grades [3], in others the difference between male and female students was very small [4–6], and in some female students did better than male students [7]. Seymour and Hewitt [39] found evidence of masculine cultures driving highly competent women out of STEM majors. While women perform as well or better than men in their STEM courses, they disproportionately left because of the hostile and competitive environment they experienced in STEM courses [39]. While Seymour and Hewitt [39] looked at STEM courses broadly, much of their data came from physics courses where most women report experiencing sexual harassment at some point during their physics education [15,16].

Growing evidence indicates that differences between male and female students in domain specific self-efficacy (i.e., physics self-efficacy) are unique to physics. Nissen and Shemwell [4] found that female and male students experienced similar self-efficacy in other STEM courses but that female students experienced much lower self-efficacy than male student only in physics courses. Henderson *et al.* [9] reviewed the literature on self-efficacy in STEM courses and found consistent gender differences in physics, but not in chemistry, mathematics, or biology. Henderson and colleagues also found that while both male and female students' self-efficacy tended to decrease in physics courses, it tended to increase in chemistry, mathematics, and biology courses.

### C. Racism in physics

Few studies have investigated race or racism in college physics [17]. Quantitative studies that have examined race or racism tend to combine students into two groups: majority (White and Asian) and underrepresented minority (Black, Indigenous, Hispanic, and all others). Some studies only look at differences after instruction while other studies control for preexisting differences. Kost *et al.* [6], Watkins [18], Brewe *et al.* [19], all used research-based assessments (e.g., the Force Concept Inventory [45]) as a pretest and a post-test for conceptual knowledge. All three studies found the differences in conceptual knowledge after instruction between majority and underrepresented minority students were explained by preexisting differences. In contrast, Van Dusen *et al.* [21] and Van Dusen and Nissen [20] found that racial differences in conceptual knowledge increased from pretest to post-test and the differences on the post-test were not explained by preexisting differences. In both cases, the college physics classroom either perpetuated or added to inequities across races.

While these quantitative studies tended to combine marginalized groups, some qualitative studies have focused on the lived experiences of students of color [22], Black students [23], Black women [24,26–28], and women of color [29] in physics. BIPOC students experience their race as a salient component of their physics education. They are often ignored and avoided by their fellow peers and faculty members, they are dissuaded from pursuing STEM degrees by faculty members, and excluded from insider knowledge needed to succeed in their education [27,29,39,46–50]. These negative experiences occurred less frequently, however, for Black men at Historically Black Colleges and Universities (HBCUs) [50].

Intersectionality studies tell us that these burdens fall disproportionately on students with multiple intersecting marginalized identities [51,52]. While Black men faced less negative experiences at HBCUs, Black women faced exclusion at HBCUs and primarily White institutions because they face marginalization from White men, White women, and Black men [50]. Hyater-Adams *et al.* [25] found Black women experienced far more instances where being recognized as a physicist pushed them away from continuing to do physics than Black men reported. For example, Black women felt unique pressures to have to prove their skills to their peers and supervisors as well as often being viewed as "too smart" by family and friends. Clancy et al. [53] found that women of color uniquely faced barriers in astronomy and planetary sciences that White women did not face.

## IV. CONCEPTUAL FRAMEWORK

Critical race theory (CRT) began in the 1970s and 1980s as a movement among a racially diverse group of U.S. legal scholars of color to address social injustices and racial oppression [54–56]. CRT explicitly assumes racism is ingrained in our institutional structures, focuses on the narratives and counternarratives of oppressed people, and identifies the importance of interest convergence between oppressed peoples and their oppressors in creating change [57,58]. Ladson-Billings [59] provides affirmative action as a poor example of interest convergence. Affirmative action is under ongoing attack as a benefit for Black, Indigenous, and people of color and is associated with primarily benefiting Black, Indigenous, and people of color. Affirmative action in higher education, however, has primarily benefited White women [60]. White women often support households with White partners and children. Because the vast majority of the benefits of affirmative action have gone to White individuals, Ladson-Billings points out that this is a poor example of a true convergence of interests.

In the intervening years, CRT has been taken up by scholars in many fields, including education [59,61]. Each of these offshoots apply the defining characteristics of CRT, such as challenging the ideas of objectivity and claims of neutrality [54], in novel contexts. To analyze and interpret our findings, we used a Quantitative CRT (QuantCrit) [38,62,63] perspective.

### A. QuantCrit

Critical research has historically used qualitative approaches to investigate the lived experiences of people from marginalized groups and the social processes that create racist, sexist, and classist power structures [62–64]. QuantCrit emerged as a quantitative perspective [38] aligned with the core principles of critical research. QuantCrit complements qualitative studies by using large-scale data to represent student outcomes in ways that reveal structural inequities that reproduce injustices [38]. A QuantCrit perspective also pushes researchers to identify where society fails to measure the outcomes for marginalized groups [65]. For example, medical research has traditionally been performed on White men, leading to a failure to identify best medical practices for White women and for BIPOC women and men [66]. Below, we describe three principles of QuantCrit [62] and the ways we strove to embody them in this investigation:

(1) *The centrality of oppression*.—We assumed that racism and sexism are complex and dynamic processes present throughout society that we must explicitly examine lest our statistical models legitimize existing inequities. Educational inequities come from hegemonic power structures creating educational and societal systems that cater to students from dominant groups. Researchers and policymakers often refer to inequities in outcomes as achievement gaps and frame these inequities from a deficit perspective in which the differences are the attributes or the fault of minoritized students and communities [67]. Rather than use this deficit perspective, we follow the recommendation of Ladson-Billings [35,67] and reframe inequities in group performance as educational debts that society owes students due to their continual marginalization. The persistence of minoritized students within the physics discipline is a testament to the strength of the undervalued cultural knowledge, skills, abilities, and contacts developed by marginalized groups [68]. Researchers can measure some aspects of educational debts society owes students with quantitative measures (e.g., representation, grades, student scores). However, quantitative measures cannot access all aspects of the educational debts owed by society, nor can a single quantitative measure indicate that an intervention or institution has redressed all educational debts.

(2) *Categories are neither natural nor given*.—All data are socially constructed and reflect the hegemonic power structures that created them. "Expertlike" views on the CLASS, for example, are social constructs created by researchers and codified by our educational systems. What expertlike views on the CLASS reflect are the expressed views of the physics faculty interviewed during the development of the instrument. The views of these physics faculty represent a particular perspective on the epistemology of physics. As the faculty were primarily employed by research-intensive institutions in the United States, it is likely that a majority of the interviewees were White men. The lack of diversity in interviewees likely led to a similar lack of diversity in perspectives on the epistemology of physics. Even within this somewhat homogeneous group of physics faculty, there was not complete agreement on what an expert view was on a number of the questions. While the CLASS gives a score for how expertlike a student's views of physics are, these scores are socially constructed and do not represent an abstract truth about a student.

Our models aggregate students by race and gender. These categories do not represent any natural or scientific truth about students but are social constructs that maintain hegemonic power structures. The dynamic socially negotiated natures of race and gender does not diminish the very real effects of racism and sexism associated with them.

We strive to clarify that our models are not measuring innate differences in students based on their race or gender, but the impacts of multidimensional oppressive power structures on students marginalized by these social constructs. One way that we reflect this in our writing is through the explicit naming of racism and sexism in interpreting our models.

(3) *Data are not neutral and cannot speak for themself.*—We reject the idea that data are neutral and can speak for themself. Hegemonic assumptions can shape every stage of collecting, analyzing, and interpreting data [69]. For example, the data we analyzed in this investigation came from the Learning About STEM Student Outcomes (LASSO) platform. While the LASSO platform has been found to be more representative than the published literature [70] and is free to instructors and students to use, it requires that faculty know of its existence, be familiar with and interested in using research-based assessments, and be willing to administer them online. Similarly, students must have access to online technologies. These barriers to participation create bias in the data, such as overrepresentation of well resourced, research-intensive institutions that often under-enroll BIPOC students.

In analyzing these data, we used methods that we felt produced the most meaningful representation of the impacts of racism and sexism knowing that the data and methods were imperfect. For example, our use of corrected Akaike information criterion (AICc) to select our models and not using $p$ values to interpret them allowed us to model and discuss inequities in student outcomes that would have been lost using more traditional methods. Other methods, however, had clear limitations. For example, there were missing data in our dataset from some students not taking either the pretest or post-test. Nissen *et al.* [71] found that the students who were most likely to participate on concept inventories were those that earned high grades in the course. We used multiple imputation [72] to create complete datasets that minimized the impact of bias from selective participation. The lack of additional strong predictors of performance (e.g., final grade), however, limited the ability of multiple imputation to account for biases in the data [73]. In creating and interpreting our models, we did our best to speak for the data in ways that identify injustices while acknowledging that our findings were shaped by our own imperfect methods.

The underrepresentation of students from marginalized groups makes it difficult to collect large enough samples to investigate inequities with dependable statistical analyses. Disaggregating across intersecting marginalized identities, such as for women of color, exacerbates these challenges. Investigations may incorrectly claim they found no differences across demographic groups because the analyses were underpowered and did not find differences with a $p$ value below 0.05. Lack of a statistically significant $p$ value should not be confused with lack of a meaningful difference or effect [74–76]. QuantCrit researchers, in part, overcome this challenge by collecting large-scale datasets with enough statistical power to model the relationships between student's intersectional identities and their learning outcomes. While our work's foundations lie in the QuantCrit literature, we were informed by the prior work using intersectionality in physics [27,29,47–50], intersectionality in QuantCrit [63,64,69], and the foundational work in intersectionality [51,52]. This body of work was particularly informative for our statistical model building process. The recent emergence of large-scale databases of university science student data [77–79] have made it easier to get the statistical power needed to model the impacts of intersecting racist and sexist power structures. Even with these large-scale databases, small samples for intersectional and underrepresented populations can obscure inequities. Rather than include $p$ values in our findings [80], we focus on transparency by reporting the point estimates and uncertainties from our models. This method, which is detailed in Sec. V E, prevents our results from focusing solely on groups well represented in the data but rather on inequities that warrant attention.

### B. Operationalizing equity

Because data cannot speak for themselves, we follow the advice of Rodriguez *et al.* [81] and Stage [38] and operationalized equity to interpret our findings from an antiracist perspective. We adopted Kendi's [82] definition of antiracism as the ideas, beliefs, and policies that hold racial groups as equal. Therefore, racial disparities result from racial discrimination. We operationalized equity as *equality of outcomes* to align with this definition. Equality of outcomes occurs when students from different gender, race, and ethnic groups have the same average achievement at the end of a course regardless of their backgrounds. This perspective on equity has been called equity of parity [81,83] and equality on average across social groups [84]. We renamed it to align with Lee's [85] definition of equity and equality. This perspective takes a strong social-justice stance as it argues a just education system must allocate resources to eliminate inequities created by discrimination. In these scenarios, students from marginalized groups receive more resources than their peers from dominant groups to counteract the effects of racist and sexist power structures and to repay the educational debts owed by society.

### C. Equity orientation

Philip and Azevedo [86] examined perspectives in the literature on informal science learning around issues of equity, diversity, and access. They summarize the equity orientations as either (i) creating new opportunities for students from historically [87] marginalized groups but not altering the status quo of what doing science in a field means or (ii) opening new possibilities for societal transformation around what it means to do science, but are less likely to impact students achievement in school directly. Philip and Azevedo [86] call on researchers to define the equity orientation their work uses.

In this article, our orientation emphasizes supporting students now over transforming what it means to do physics and who gets included in physics. We emphasize supporting students now because we feel the subset of attitudes we focus on are good outcomes in physics courses and individual instructors can enact changes in their classrooms to make these changes happen now. The scientific literature contains multiple examples of pedagogies individual instructors can use in their courses to support students in developing the attitudes they need to succeed in physics [8]. Using pedagogies that support attitude development of students from historically marginalized groups can create interest convergences because they may also improve overall recruitment and retention of physics majors in their department. Many cultural attitudes in physics (e.g., competition, individualism, and solitary practice) are more costly for women and BIPOC students to adopt [29,39]. Our focus, however, is on pedagogies that support students in seeing physics as applicable to their lives, as understanding that physics is more than plugging numbers into the right equations, and feeling capable of learning and doing physics. We pursue these goals in our own courses. By enacting these changes now, we expand the foundation for redefining what it means to learn and do physics in two ways. First, we support physics educators in reflecting on and changing their pedagogical practices. Second, we support more students from historically marginalized groups becoming physicists. These two groups of physicists may support the broader physics community in transforming what it means to learn and do physics to create a more inclusive culture in physics.

### D. Positionality

Feminist theory has shown that all knowledge is marked by those who create it [88]. To be transparent about the position of the researchers in this work in relation to the power structures under investigation, we offer positionality statements [65] for each of the authors.

The following is the first author's, J. N., positionality statement. My identity as a White, cisgendered, heterosexual, nondisabled man has provided me with power and opportunities denied to others in American society. I use my experience growing up in a poor home and as a veteran of the all-male submarine service to motivate reflecting on and working to dismantle my privilege. My work on this project was shaped by the post-positivist scientific traditions I was educated in and my activist goal to pursue scientific knowledge that can help identify and dismantle policies and systems of oppression. Because of the privilege implicit in my current identities, I brought a limited perspective to this work on racism and sexism.

The following is the second author's, I. Her Many Horses, positionality statement. I identify as a Lakota (Indigenous), cisgender, heterosexual, man and was raised on the Rosebud Reservation in South Dakota. I consider myself to be educationally privileged and am a third generation college student with many family members holding terminal degrees. I hold an undergraduate degree in computer science and a Ph.D. in education. Throughout my life I am usually the only person that looks like me anywhere I go. These experiences have driven me to use my own power to address issues of equity in whatever space I find myself.

The following is the third author's, B. V D., positionality statement. I identify as a White, cisgender, heterosexual, man with a color vision deficiency. I was raised in a pair of lower-income households but I now earn an upper-middle class income. I hold an undergraduate degree in physics and a Ph.D. in education. I am an assistant professor at a Hispanic serving institution. My experiences working with marginalized students, particularly those whom I have had the honor to mentor as learning assistants [89] and as researchers, has motivated my attempts to use my position and privilege to dismantle oppressive power structures. As someone who seeks to be an ally, it is easy to overlook my own privileges. I try to broaden my perspective through feedback from those with more diverse lived experiences than my own.

## V. METHODS

### A. Instrument

We used data collected with the Colorado Learning Attitudes about Science Survey [11] for this study. Researchers and educators commonly measure attitudes in college physics courses using either the CLASS or the Maryland Physics Expectations Survey (MPEX) [90]. The surveys ask students about several different categories of attitudes about physics, such as the relationship between learning physics and everyday life, the effort they put into learning physics, and their approach to solving physics problems. Students respond to these questions on a five-point Likert scale from strongly agree to strongly disagree. Researchers use the instruments to create an overall score and a score for each of the categories of attitudes based on how many times the students agree or strongly agree with what expert physicists reported [11,40].

The CLASS is the most commonly used measure of attitudes and attitudes about learning and doing physics [33]. However, some researchers have raised questions about what the CLASS measures and how the CLASS is analyzed [40,91–93]. The original authors of the CLASS argued that it measured a set of eight categories of attitudes. However, three research teams [40,93–95] used common psychometric methods to find a smaller set of factors measured by the CLASS. Table I provides example questions for each of the factors Douglas *et al.* [40] described: personal application, personal effort, and problem solving. This structure is similar to the one found by Kontro and Buschhüter [93] and by Heredia and Lewis [92] on the CLASS for chemistry. The original authors also adopted the reduction of the scores into two categories with agree and strongly agree categorized as expertlike and the other three options categorized as not expertlike. However, Van Dusen and Nissen [91] found evidence against collapsing the responses into two categories. They concluded that this practice unnecessarily discards unique information.

TABLE I. Examples of statements the CLASS gives students to rate their level of agreement with.

| Personal application |
|---|
| I think about the physics I experience in everyday life. |
| I study physics to learn knowledge that will be useful in my life outside of school. |
| To understand physics, I sometimes think about my personal experiences and relate them to the topic being analyzed. |

| Personal effort |
|---|
| In doing a physics problem, if my calculation gives a result very different from what I'd expect, I'd trust the calculation rather than going back through the problem. |
| In physics, it is important for me to make sense out of formulas before I can use them correctly. |
| To learn physics, I only need to memorize solutions to sample problems. |

| Problem solving |
|---|
| After I study a topic in physics and feel that I understand it, I have difficulty solving problems on the same topic. |
| If I want to apply a method used for solving one physics problem to another problem, the problems must involve very similar situations. |
| If I get stuck on a physics problem, there is no chance I'll figure it out on my own. |

In addition to these concerns about how the CLASS data are analyzed, we are also concerned about who was used to determine what expertlike attitudes in physics are. Adams *et al.* [11] describes interviews with three expert physicists to establish the validity of the questions. Sixteen physicists with extensive experience teaching introductory courses or interest in teaching then took the survey to confirm the scoring of the items. But, the paper does not provide details on the institutions or demographics of these expert physicists, and three physicists feels insufficient to establish the expertlike attitudes physics education should focus on. We are particularly concerned that the development of the CLASS did not take into account the attitudes from a broad array of physicists with diverse backgrounds at diverse institutions in a variety of roles. International researchers translating the CLASS into eight languages [96–99] besides English, however, indicates physicists from around the world value the attitudes measured by the CLASS.

While the CLASS is not perfect, it is widely used by both researchers and educators. It has played an important role in supporting the physics education research community in focusing on diverse factors critical to student success. Therefore, we concluded the CLASS could provide a useful data source to investigate equity in students' attitudes about learning and doing physics.

### B. Data collection, cleaning, and imputation

We accessed student and course data through the Learning About STEM Student Outcomes platform [100]. The LASSO platform administers, scores, and analyzes research-based assessments online to build a multi-institution database. This database of student and course data is anonymized and filtered to only include students that consented to share their data. Educators using the LASSO platform tended to administer the CLASS twice: as a pretest during the first week of class and as a post-test during the last week of class. The analyzed data came from 4673 students in 95 first semester introductory college physics courses: 2503 students in 49 algebra-based courses and 2170 students in 46 calculus-based courses. The courses were taught at 18 institutions including 2 two-year colleges and 2 private universities.

We differentiated between calculus-based and algebra-based physics courses because these courses typically serve two different sets of majors and the enrolled students often identify with different demographic groups. Kanim and Cid [101] point out that little work in physics education research examines algebra-based courses and this aligns with the review of research on attitudes in physics by Madsen *et al.* [33]. Algebra-based physics courses serve nonphysical science majors, such as various biology majors and at some institutions serve engineering technology majors. Algebra-based physics courses that serve biology majors tend to have similar proportions of students that identify as men and women whereas men tend to make up the vast majority of students in courses that serve engineering technology majors. The courses serving biology majors, to our knowledge, are much more common than courses serving engineering technology majors. Calculus-based physics courses primarily serve physical science, engineering, and mathematics majors. These courses tend to have approximately 80% men and 20% women. We are not aware of any consistent differences in the race and ethnicity of the students in these courses. While few studies have investigated algebra-based courses [33,101], results in this study show CLASS scores differ between these two course types.

We scored the CLASS responses using the agree categories recommended by the original authors and we only analyzed the total score of the 36 items they include in their total scores [11]. We followed their original scoring recommendations so our results would be comparable to prior research using the CLASS. Adams *et al.* [11] recommends not including 6 of the 42 items in the total score. One excluded item is a filter question. Experts did not consistently agree on four of these items. Two items ask about the nature of science and two others ask about learning styles. The final excluded item also asks about approaches to learning but is not discussed by Adams *et al.* [11]. These excluded items and the extensive process Adams *et al.* [11] details illustrate that expert's attitudes vary.

To clean the data, we removed the pretest or post-test score if the student took less than 3 min on the assessment or incorrectly answered the filter question [11]. We removed any courses with less than 5 pretests or 5 post-tests. After cleaning the data, we used hierarchical multiple imputation (HMI) with the hmi [102] and mice [103] packages in RStudio V. 1.1.456 to impute missing data. We only imputed values for missing pretest and post-test CLASS scores, and we did not impute missing values for gender and race to respect each student's choice to not answer these questions. HMI provided a principled method for handling missing data that maximized statistical power and minimized bias while accounting for the hierarchical structure of the data [73,104–107].

The imputed dataset included 7764 students. This imputed dataset was larger than the 4673 students used in the analysis because it included students enrolled in a variety of courses: first and second semester algebra and calculus based physics courses, LA pedagogy courses, upper division physics courses, and physics courses for education majors. The rate of missing data for this dataset was 17% on the pretest and 34% on the post-test. The imputation model included a dependent variable for the post-test and accounted for the pretest score, course type, and demographic variables and nested the students within courses. The subsequent analysis only included 4673 students enrolled in first-semester algebra-based and calculus-based introductory physics courses.

### C. Model building

To investigate student attitudes, we developed models to predict student attitudes on the pretest and post-test and in algebra-based and calculus-based first-semester physics courses separately, which are described by $CLASS_{ij}$ in the final model. The models were 2-level hierarchical linear models with student data in the first level and course data in the second level. Using hierarchical linear models accounted for the nested nature of the data [108,109]. We ran the models and pooled the results for the imputed datasets using the mitml [110] and lme4 [111] packages in R. The hierarchical linear model parameters were fit using the penalized least squares method.

Woltman *et al.* [112][p. 56] provides a detailed description of HLM equations, which we will cover briefly here. The subscripts for $CLASS_{ij}$ refer to the $i$th student in the $j$th course. The $\beta_{0j}$ is the intercept for the $j$th course. The $\beta_{(1-12)j}$ represents the slope (e.g., the regression coefficient) for each variable for the $j$th course. The $r_{ij}$ term represents the student-level error associated with the $i$th student in the $j$th course. The $r_{ij}$ term is analogous to the $\epsilon$ term in standard linear regressions, which is often omitted in representations of the equations for linear regressions. It represents the difference between the predicted and actual values. In the level-2 equations, the $\gamma_{00}$ represents the overall mean intercept. The $\mu_{0j}$ term represents the course-level error associated with the $j$th course and is what allows the intercept to vary across each course. The model is a fixed slope model since the slopes, $\beta_{(1-12)j}$ equations do not include a $\mu$ variable.

The dataset included demographic data for gender, race, and ethnicity. The gender identity question included options for male, female, genderqueer or gender nonconforming, trans female or trans woman, trans male or trans man, another identity with the option to write in a description, and prefer not to answer. Students could select all answers that applied to them. These response categories conflate gender and sex, which may have shaped how some students responded to these questions. To address this mislabeling of student gender, we will use the terms men and women rather than male and female. The ethnicity question had options for Hispanic or Latino, not Hispanic or Latino, and prefer not to answer and students could select one. The race question had options for White, Black or African American, Asian, American Indian or Native American, Native Hawaiian or other Pacific Islander, a race not listed with a write in option, and prefer not to answer. Students were able to select multiple responses for the race question.

#### 1. Final model

Level-1 equations (Student level)

$$\begin{aligned} CLASS_{ij} = &\ \beta_{0j} + \beta_{1j} \times \text{Gender other} + \beta_{2j} \times \text{Hisp.} + \beta_{3j} \\ &\ \times \text{White} + \beta_{4j} \times \text{Woman} + \beta_{5j} \times \text{Black} + \beta_{6j} \\ &\ \times \text{Asian} + \beta_{7j} \times \text{Race Other} + \beta_{8j} \times \text{Hisp.} \\ &\ \times \text{White} + \beta_{9j} \times \text{Woman} \times \text{Black} + \beta_{10j} \\ &\ \times \text{Woman} \times \text{Asian} + \beta_{11j} \times \text{Woman} \times \text{Hisp.} \\ &\ + \beta_{12j} \times \text{Woman} \times \text{White} + r_{ij}. \end{aligned}$$

Level-2 equations (Course level)

$$\beta_{0j} = \gamma_{00} + \mu_{0j},$$
$$\beta_{(1-12)j} = \gamma_{(1-12)0}.$$

To determine what demographic variables to include in the models, we first used a rule of thumb to only investigate scores for populations with at least 20 students total [113]. This meant that we did not include variables for transgender, Hawaiian or Pacific Islander, or Native American in our models. Because removing the students with these identities could have biased the course-level results and because some students did not include a gender or race, we combined these students into two categories: gender other and race other. This meant that the final variables used in our model, which is shown above, included woman, gender other, Black, Asian, Hispanic, White, and race other. We included interactions between variables whenever a population included more than 20 students but not for the race other and gender other groups. Hispanic is often treated as an ethnicity in the United States [114]. However, 67% of Hispanic Americans consider their Hispanic identity to be a part of their racial identity [114]. We found a similar trend in our data to those described by Parker *et al.* [114] where many Hispanic students either selected "a race not listed" or did not choose a race. Therefore, we treated Hispanic as a racial identity and we did not include these students in the race other category or interact Hispanic with the race other group in our models. The data included sufficient students who identified as White and Hispanic to include an interaction term in our model, but it included too few students who identified as Black and Hispanic or Asian and Hispanic to include those interaction terms in the model. We interacted woman with each of the racial groups in the model.

We wanted to build as simple of a model as possible that reflected the student's identities. This parsimony protected against including unnecessary variables that reduced the degrees of freedom [115]. The reduction in degrees of freedom would have increased the standard errors for all of the coefficients. However, we needed to balance parsimony with the accuracy of our representation of student identities in our models. Specifically, we wanted to balance accounting for intersectionality through interactions between variables with the concern that including interactions that were not present in the data decreased the models' ability to identify other relationships.

To support parsimony in our model building, we used the dredge package in R [116] to determine if a simpler model than our initial one provided a better fit [117]. The dredge package took the data and our initial model and analyzed the initial model and every simpler iteration of that model using the Akaike information criterion corrected (AICc). A simpler iteration removed one or more variables or interactions. We used the differences in AICc scores between models to identify the quality of the fit with a rule that any model with a AICc score that was at least 2 higher than the lowest score was not a good fit [118]. Dredging the models indicated that our initial model provided the best fit for both the algebra-based and calculus-based course data.

To inform the educational significance of the educational debts society owes students due to racism and sexism we identified, we drew on the literature to establish a cutoff in expertlike attitudes necessary for students to pursue a graduate degree and become a professional physicist. Based on Gire *et al.* [30] and Bates *et al.* [31] professional physicists score approximately 85% expertlike views on the CLASS on average with a standard deviation of approximately 10%. Given that students do not tend to improve their attitudes about physics through their undergraduate education [30,31], we used 75% as an estimate for the attitudes a student needs to start college with to have a reasonable chance of becoming a professional physicist. This 75% cutoff is not definitive for the minimum attitudes to become a physicist. Rather, we use it to illustrate the potential effects of educational debts in physics attitudes that society owes students due to racism and sexism.

We built hierarchical generalized linear models (HGLM) of the probability that students scores were equal to or greater than 75%. We developed our HGLM models using the hglm package in R [119] and the parameters were fit using the extended quasilikelihood method. These models had identical independent variables to those in our hierarchical linear model. We focus on the models for pretest scores greater than or equal to 75% because the pretest scores had less missing data, they represent a best case scenario since scores tended to decrease, and the posttest scores in algebra-based physics courses had so few students from marginalized groups score greater than or equal to 75% that the models failed to converge.

### D. Descriptive statistics

We provide the full descriptive statistics in the Appendix along with a figure illustrating the distribution of scores within each demographic group. Table II shows the sample sizes for each demographic group.

TABLE II. Sample sizes for each demographic group. The "other gender" group included 29 students in algebra-based and 33 students in calculus-based courses.

|  | Calculus-based | | Algebra-based | |
| --- | --- | --- | --- | --- |
|  | Women | Men | Women | Men |
| All | 697 | 1444 | 1462 | 1008 |
| Asian | 154 | 270 | 309 | 166 |
| Asian Hispanic | 4 | 7 | 5 | 4 |
| Black | 42 | 63 | 98 | 40 |
| Black Hispanic | 7 | 5 | 8 | 3 |
| Hispanic | 43 | 114 | 65 | 58 |
| Other | 55 | 105 | 96 | 71 |
| White | 344 | 726 | 760 | 583 |
| White Hispanic | 48 | 154 | 121 | 83 |

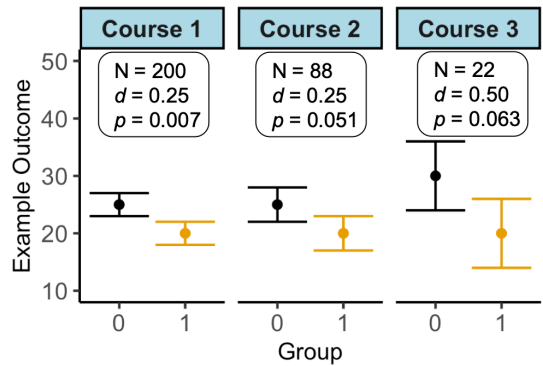

FIG. 1. An example dataset comparing two groups in three different courses. Course 1 had a small statistically significant difference. Course 2 had the same size difference as course 1 but it was not statistically significant. Course 3 had a difference twice as large as course 1 but it was not statistically significant. The different course enrollments ($N$) across these three courses shows how sample size impacts the uncertainty in the differences but not the size and meaningfulness of the differences. $N$ is the course enrollment, $d$ is the effect size, $p$ is the $p$ value, and the error bars represent 1 standard error.

### E. Interpreting results

We do not present $p$ values. $p$ values depend on sample size and lead to selective reporting and selective attention [80] that can ignore injustices borne by the most underrepresented and marginalized groups of students. Our analysis, instead, focused on the point estimates and standard errors produced by the models. This decision was informed by our QuantCrit perspective, which pushed us to question common statistical practices, and aligns with recommendation from the American Statistical Association in response to scientists and scientific communities misuses of $p$ values [74,76,80].

Instead of using $p$ values, we used the overlap in the standard errors of the point estimates to inform our confidence in the results by indicating how compatible the model was with the data. An overlap between one standard error bars approximately produces a $p$ value of 0.05 for a one-sided $t$ test. We did not, however, use overlap as a binary indicator of significance. Figure 1 provides an example dataset illustrating three courses. Course 1 has a small statistically significant difference. Course 2 had the same sized difference as course 1 but it was not statistically significant because the data came from fewer students. The difference in course 3 was twice as large as that in courses 1 and 2 but it was not statistically significant because of the even smaller enrollment in the course. Differences in the size of the error bars often result from differences in sample sizes. $P$ values combine the size and uncertainty in the differences into one measure, and therefore obscures the distinction between the two. This practice is particularly problematic in equity research where the focus is often on historically excluded groups with low levels of representation. In our example data, if one uses $p$ values as binary indicators of significance the differences in course 3 would be rejected as not statistically significant and potentially not even reported though they were twice as large as the differences in course 1 and 2.

To better inform our interpretation of model uncertainty, we also used the consistency of results across similar comparisons. In our example data, Fig. 1, all three courses tell the same story about the differences between the groups. That consistency increases our confidence in those differences. In the analysis, we used this approach when looking at differences between women and men for each racial group. Small sample sizes within many of the racial groups, Table II, caused relatively large standard errors for those measurements. In daily life this is similar to rolling a pair of six-sided dice. If you role them once and get twelve that is not remarkable. If you role them six times and get twelve each time, you can be pretty confident that the dice are not random.

## VI. FINDINGS

The first findings section covers the predicted mean expertlike attitudes across demographic groups for algebra-based and calculus-based courses. The second findings section covers the models predicting the proportion of students from each demographic group with greater than the 75% cutoff for expertlike attitudes. The sections each discuss the models in terms of educational debts due to sexism, then racism, and then the intersectionality of racism and sexism. Within each of these discussions, we first discuss the sizes of the differences and then the uncertainty of those differences. We used the 75% cutoff models to interpret the practical significance of the educational debts on students' opportunity to become physicists. The predicted means models covered both pretest and post-test scores while the proportion greater than the 75% cutoff model only focused on pretests due to the issues of missing data discussed in the methods, Sec. V C. In both findings sections, we provide the predicted values for each demographic group based on these models as both a figure and a table. The figures provide visual comparisons across all groups and trends in the data while the tables provide precise values for specific comparisons and for use by future researchers. To simplify the findings section and provide transparency and utility for future researchers, we provide the model outputs in the Appendix.

### A. Predicted attitudes

The mean predicted attitudes across demographic groups indicated students tended to have more expertlike attitudes in calculus-based physics courses than in algebra-based physics courses. The predicted values also showed attitudes decreased for all ten demographic groups in the calculus-based physics courses; see Table III. This decrease in attitudes in the calculus-based courses varied across demographic groups and ranged from −0.1 to −5.7 percentage points. The changes in attitudes covered a similar-sized range in the algebra-based courses but was more centered around zero with decreases for just six of the ten demographic groups and ranged from 1.8 to −2.5 percentage points. These small increases to moderate decreases in attitudes align with findings of near zero change to small decreases in attitudes in most introductory physics courses that don't use a pedagogy designed to support attitude development [33].

TABLE III. Predicted values for average attitudes based on the hierarchical linear models. The table includes the point estimate (Est.) and its standard error (SE).

| | | Algebra-based | | | | Calculus-based | | | |
| | | Pre | | Post | | Pre | | Post | |
| Race | Gender | Est. | SE | Est. | SE | Est. | SE | Est. | SE |
| --- | --- | --- | --- | --- | --- | --- | --- | --- | --- |
| Asian | Women | 53.6 | 1.1 | 53.2 | 1.3 | 56.2 | 1.4 | 56.1 | 1.7 |
| | Men | 57.4 | 1.4 | 54.9 | 1.6 | 60.4 | 1.2 | 59.1 | 1.4 |
| Black | Women | 51.3 | 1.8 | 48.8 | 2.4 | 57.8 | 2.4 | 52.2 | 2.9 |
| | Men | 54.8 | 2.7 | 55.5 | 3.3 | 61.4 | 2.2 | 56.5 | 2.6 |
| Hispanic | Women | 53.9 | 1.9 | 55.6 | 2.2 | 56.2 | 2.2 | 54.1 | 2.6 |
| | Men | 55.5 | 2.1 | 54.7 | 2.4 | 58.6 | 1.6 | 55.1 | 2.0 |
| White | Women | 56.3 | 0.7 | 55.0 | 0.9 | 63.9 | 1.0 | 60.9 | 1.3 |
| | Men | 59.4 | 0.8 | 59.8 | 1.0 | 67.1 | 0.8 | 64.7 | 1.1 |
| White Hispanic | Women | 53.8 | 1.5 | 53.9 | 1.7 | 60.0 | 2.0 | 57.9 | 2.6 |
| | Men | 55.7 | 1.8 | 56.0 | 2.2 | 64.6 | 1.3 | 61.2 | 1.8 |

We interpreted the consistent gender differences within races in both course types as indicating the consistent educational debts society owes women based on our conceptual framework, as shown in Fig. 2 and Table III.

In the calculus-based courses, men had more expertlike attitudes than women for all racial groups. These educational debts owed by society ranged from 2.4 to 4.6 percentage points on the pretest and 1.0 to 4.3 percentage points on the post-test. Society's educational debts in attitudes for women in the algebra-based courses varied from 1.9 to 3.8 percentage points on the pretest and −0.9 to 6.7 percentage points on the post-test. The one exception for society's consistent educational debts was to Hispanic women who had more expertlike attitudes than Hispanic men on average at the end of algebra-based courses.

To inform our confidence in the size of society's educational debts due to sexism, we looked at the overlap in the error bars between women and men's predicted attitudes within races in addition to the consistency of the educational debt owed due to sexism across races. Figure 2 shows the error bars overlap for 6 of the 10 comparisons of men and women within races in both algebra-based and calculus-based courses. This overlap indicates we cannot be overly confident about the size of society's educational debts owed to women for some races nor about how this educational debt in attitudes changed from pre- to post-instruction. This lack of confidence resulted from the differences in predicted average scores between men and women being small in some cases and because the low representation for BIPOC students in the data created larger uncertainties around the predicted average scores for those demographic groups. However, the estimated average attitudes indicated society owed educational debts to women for 19 of the 20 comparisons (5 races across 2 course types at 2 times). The consistency of this result indicated that society has provided women, in general, inferior opportunities to develop expertlike attitudes about physics than men in college physics courses.

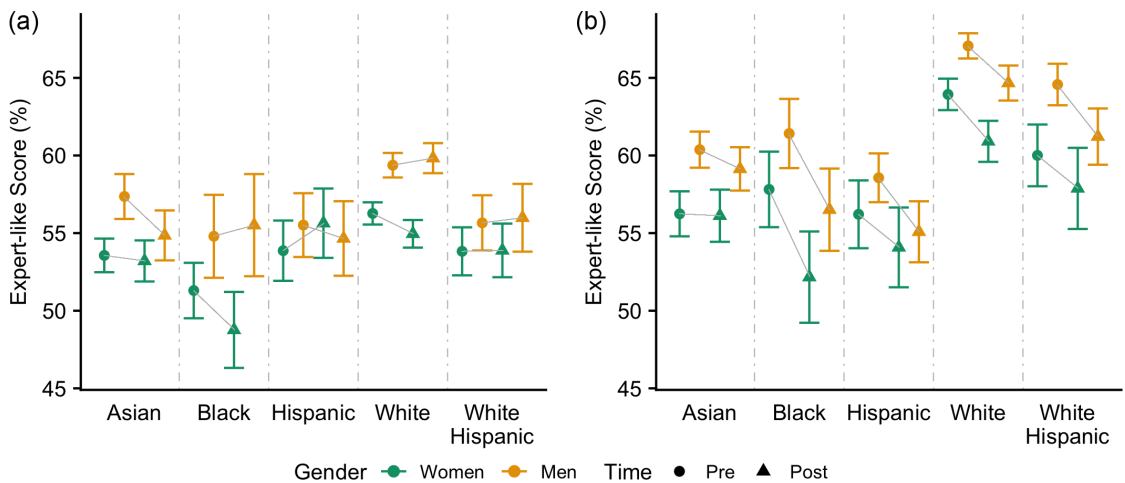

FIG. 2. Predicted pretest and post-test attitudes as measured by the CLASS for (a) algebra-based and (b) calculus-based courses. The point estimates for algebra-based courses show lower scores overall than calculus-based courses. The shifts from pretest to post-test varied but tended to be small; attitudes tended to become less expertlike for all students in calculus-based courses but shifts varied across groups in algebra-based courses. The differences in attitudes tended to favor men and White students but these trends were not consistent and many of the differences were small. For calculus-based courses, the point estimates show large differences in attitudes favoring male and White students and show decreases for most groups of students. Error bars represent 1 standard error.

Based on our conceptual frameworks, we interpreted the predicted values for students in both course types showing higher scores for White students than for Asian, Black, Hispanic, or White Hispanic students as representing an educational debt owed by society to BIPOC students. This educational debt owed by society was twice as large on average in calculus-based courses. Comparing White students to BIPOC students who were the same genders, in algebra-based courses the differences ranged from 2.0 to 4.6 percentage points higher on the pretest for White students and the range increased to vary from −0.6 to 6.2 percentage points on the post-test. In the calculus-based physics courses the differences in attitudes between White students and BIPOC students ranged from 2.5 to 8.5 percentage points on the pretest and from 3.0 to 9.6 percentage points on the post-test.

Investigating the overlap in the error bars informed how certain we were about the existence of society's educational debts due to racism. In both course types, all but one of the differences in average attitudes between White men and men of color were larger than the uncertainty in the two estimated means showing that this data strongly indicated society owed educational debts to men of color. For White men in algebra-based courses on the pretest the minimum value of the error bar was 58.6%. This overlapped by a small amount with the maximum value of the error bar for Asian men 58.8%, but did not overlap with the values for Black (57.5%), Hispanic (57.6%), and White Hispanic men (57.5%). A similar lack of overlap occurred on the pretest in the calculus-based courses with a minimum value of the error bar for White men (66.4%) exceeding the maximum values for Asian (61.6%), Black (63.6%), Hispanic (60.2%), and White Hispanic men (65.9%). Similar patterns existed on the post-test and can be calculated from Table III. This measure of society's educational debt due to racism was also larger than our uncertainty in the estimated average attitudes for women in the calculus-based courses. The minimum value of the error bar for White women in calculus-based courses on the pretest (62.9%) exceeded the maximum value of the error bar for Asian (57.6%), Black (60.2%), Hispanic (58.4%), and White Hispanic women (65.9%). This trend was similar on the post-test except the error bars overlapped for White and White Hispanic women. In algebra-based courses, the differences in the predicted average attitudes for women were larger than the uncertainty of the measurement in most cases on the pretest but not on the post-test. The minimum value in the error bar for White women on the pretest (55.6%) exceeded the maximum value for Asian (54.7%), Black (53.1%), and White Hispanic women (55.3%) and had a small overlap with Hispanic women (55.8%). This difference in overlap of the error bars between the pretest and post-test occurred because the greater rate of missing data on the post-test increased the uncertainty in those estimated average attitudes and the post-test average attitudes for White women decreased more than for all other groups of women except Black women. Overall these results point to society owing a consistent educational debt due to racism. The lower certainty for society's educational debt due to racism for women at the end of algebra-based courses illustrated how racism and sexism interact differently in different contexts.

Comparing White students and all other races showed consistent differences in attitudes representing society's educational debts due to racism. Looking across Black, Hispanic, White Hispanic, and Asian students, the differences in predicted average attitudes between these groups were small compared to the uncertainties in the measurement in most cases. This data did not indicate society owed students from one race a larger educational debt than another. The one exception to this trend in the data is for White Hispanic students. In the calculus-based courses, the attitudes of White Hispanic students were more similar to White students than to Hispanic students. This relationship was present but very small in the algebra-based courses.

Researchers often combine Asian students with White students because both groups are overrepresented in physics. The results in both algebra- and calculus-based courses show that society owed Asian students a similar educational debt to other BIPOC students.

Society bears a double debt to women of color due to educational debts resulting from both sexism and racism. This double burden was illustrated by White men having the most expertlike average attitudes and Asian, Hispanic, and Black women tending to have the least expertlike average attitudes. The models associated more expertlike attitudes with calculus-based courses, being White, and being a man. In the calculus-based courses, society owed women of color a greater educational debt than White women or men of color. This relationship was less clear in the algebra-based courses where many of the differences between White women and women of color were smaller than the uncertainty in the measurements. However, in both courses society owed the greatest educational debt to Black women whose predicted average attitudes were 8.1 to 12.5 percentage points lower than the predicted average attitudes for White men. To interpret the size of these differences in the context of becoming a physics major, we explore the proportion of students from each race and gender above the 75% threshold of attitudes held by most physicists in the next section.

TABLE IV. Predicted values for the proportion of students who scored above the 75% threshold on the pretest based on the hierarchical generalized linear models.

| Race | Gender | Algebra Est. [+/− 1 SE] | Calculus Est. [+/− 1 SE] |
|---|---|---|---|
| Asian | Women | 0.11 [0.09, 0.13] | 0.13 [0.09, 0.17] |
| | Men | 0.18 [0.15, 0.22] | 0.21 [0.19, 0.24] |
| Black | Women | 0.07 [0.05, 0.10] | 0.09 [0.06, 0.15] |
| | Men | 0.20 [0.14, 0.27] | 0.22 [0.17, 0.28] |
| Hispanic | Women | 0.08 [0.05, 0.11] | 0.14 [0.10, 0.19] |
| | Men | 0.10 [0.07, 0.15] | 0.13 [0.10, 0.16] |
| White | Women | 0.17 [0.16, 0.19] | 0.28 [0.25, 0.32] |
| | Men | 0.20 [0.18, 0.22] | 0.36 [0.34, 0.39] |
| White Hispanic | Women | 0.09 [0.07, 0.12] | 0.19 [0.15, 0.25] |
| | Men | 0.16 [0.12, 0.20] | 0.27 [0.24, 0.31] |

### B. Proportion of students above 75%

As we described in the methods section, Table IV and Fig. 3 represent the results of the hierarchical generalized linear models predicting the proportion of students from each demographic group who scored above 75% on the pretest. Seventy-five percent provided an estimated cutoff for the attitudes students need prior to taking their first college physics course to have a reasonable chance of becoming a professional physicist.

The results of the hierarchical generalized linear models showed society owed educational debts to women and BIPOC students. The proportion of students making the 75% cutoff ranged from a low of 7% for Black women in algebra-based courses to a high of 36% for White men in calculus-based courses. Across all groups except Hispanic students in calculus-based courses the models predicted men to be above the threshold more often than women. In 6 of 10 comparisons, these raw differences were large. In both course types, we measured a 13 percentage point gender difference for Black students and an 8 percentage point gender difference for Asian students. In calculus-based physics courses we measured an 8 percentage point gender difference for White students and White Hispanic students. These gender differences meant Black men were more than twice as likely to be above the threshold than Black women (7% versus 20%). In the two cases where the absolute difference was smaller, for example, a 3 percentage point difference for White students in algebra-based courses, the relative difference was still large. White men were 1.2 times as likely to be above the threshold than White women in algebra-based physics courses. Most of these gender differences were much larger than the uncertainties in the measurement and the consistent gender difference across 9 of the 10 comparisons indicated society owed educational debts to women whereby men are 20% to 290% more likely to meet the threshold of attitudes that physics programs filter students for. For Hispanic students, the gender differences were much smaller than the uncertainty in the measurements. However, Fig. 3 shows that this smaller difference was due largely to a much lower proportion of Hispanic men above the cutoff than for Asian, Black, or White Hispanic students. More studies are needed to determine if this gender difference for Hispanic students is reliably close to zero and to understand why that is the case.

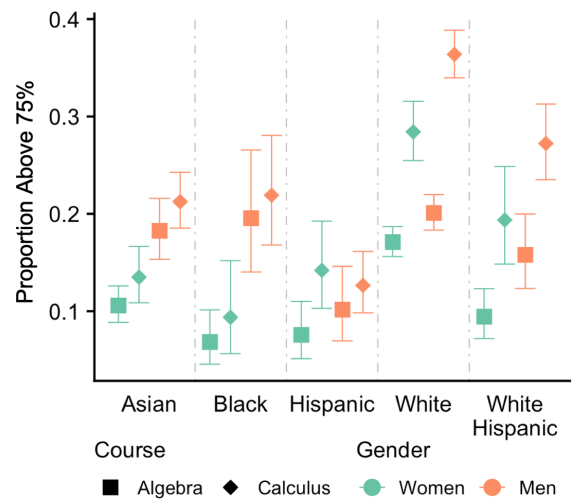

FIG. 3. Proportion of students who scored above the 75% cutoff on the pretest. Error bars represent 1 standard error.

The differences between calculus- and algebra-based courses adds complexity to interpreting the results across races. For Asian, Black, and Hispanic students, the differences between the two course types were relatively small (2–6 percentage points). In contrast, the differences across course types were relatively large for White students (11 and 16 percentage points) and White Hispanic students (10 and 11 percentage points). This difference revealed that society's educational debts owed to BIPOC students were most pronounced in calculus-based courses. The ratio of the rates between BIPOC students and White students provides a measure of the size of that difference. For women this ratio ranged from a high of White women being 3.1 times more likely than Black women to be above the 75% cutoff to a low of 1.5 times more likely than White Hispanic women. For men this ratio ranged from a high of White men being 2.8 times more likely than Hispanic men to be above the 75% cutoff to a low of 1.3 times more likely than White Hispanic men. Society's educational debts were even more pronounced from an intersectional perspective. White men were 4 times more likely than Black women to be above the 75% cutoff and this ratio ranged to a low of 1.9 times more likely than White Hispanic women.

These results point to several trends. Clear gender and racial differences indicated society owed educational debts to women, BIPOC students, and especially women of color that harmed their chances of becoming professional physicists. Society's educational debts were also owed to Asian students who researchers often combine with White students because both groups are represented in physics at higher rates than in the U.S. population. Society's educational debts were smaller for White-Hispanic students than for other BIPOC students. Lastly, the differences in attitudes between algebra-based and calculus-based courses were small for Asian, Black, and Hispanic students.

## VII. DISCUSSION

The findings showed society owed large educational debts to women, Asian, Black, Hispanic and White Hispanic students in terms of attitudes about learning and doing physics. This does not mean that society does not owe similar educational debts to students from other marginalized groups, such as Indigenous students. Rather those educational debts were likely substantial enough to result in very few students from those groups being included in the dataset. The educational debt society owed women is consistent with a lack of early educational experiences in the sciences playing a role in the underrepresentation of women in physics [10]. We are not familiar with similar work on racism, but it seems likely that society produces similar inequities in their early education for students from all marginalized groups. Though the causes of these inequities likely varies. These differences in physics attitudes do not represent deficiencies in the students themselves; rather, they represent a lifetime of science and math experiences that catered to White men. By failing to address the educational debt society owed to both women and BIPOC students, physics education perpetuates the racist and sexist power structures that created those debts. Instruction that supports students in developing expertlike attitudes in physics addresses society's educational debts and has the potential to increase the number of students from marginalized groups in physics.

Society's educational debts owed to Asian students shown in the results indicate that the common practice of grouping Asian and White students is problematic. Combining these two groups may be appropriate in situations where the data indicates they have similar outcomes and disaggregation undermines the ability to make meaningful claims. Combining them without empirical evidence, however, can obscure inequities between White students and BIPOC students. Researchers should make decisions about combining and disaggregating groups in their analysis based upon the empirical evidence within their data. Researchers can provide transparency and evidence for their decisions by providing descriptive statistics disaggregated by race and gender. These descriptive statistics will also be useful for future research and meta analyses.

Asian Americans are a panethnic group. The timings and geopolitical forces behind different emigrations and the cultural values vary across groups [64]. Using the higher representation of Asian Americans in physics as an excuse to ignore the educational and societal debts owed to Asian Americans perpetuates racist discrimination. This common practice in physics enacts the model minority myth created by anti-Black racists to perpetuate racism in the United States [120]. Researchers and institutions can begin to move away from this racist practice by collecting more fine grained information about Asian ethnicities. While researchers may need to aggregate these students to maintain statistical power, building more detailed datasets will enable future quantitative work that can break down the racist veil that hides society's educational debts owed to Asian Americans.

Algebra-based physics courses enroll diverse and talented students capable of becoming physics majors. Changes to the way those courses are taught and pathways from those courses into physics majors can improve the experiences and outcomes of the students in those courses and the recruitment of physics majors. The differences in attitudes between BIPOC students in algebra-based and calculus-based courses were relatively small. Whitten *et al.* [121] and colleagues investigated physics departments at 9 colleges to identify practices that supported recruiting and graduating physics majors with diverse identities. They found that algebra-based and teacher preparation courses were an important source of students from diverse backgrounds. While calculus-based courses were similar across campuses, algebra-based and teacher preparation courses provided faculty with opportunities to innovate and share their passion for teaching and physics. They found that the faculty at the historically Black colleges they visited were the only ones to consistently discern students who are interested and talented in physics from those who happen to have a strong physics background. These faculty also tended to be dedicated to helping students succeed in physics by addressing the educational debts society owed these students. Our data show that algebra-based courses have a small but diverse group of students with the attitudes to become professional physicists. Creating more pathways into physics supports more and more diverse students in finding their passion and pursuing their curiosity.

To provide context for interpreting differences in CLASS scores, readers can use the findings from this study to relate differences in CLASS scores to differences in the proportions of students who scored above 75%. They could also standardize the differences in CLASS scores using standard deviations in descriptive statistics to get a measure similar to Cohen's $d$. Readers could then use rules of thumb for Cohen's $d$ to interpret the practical significance of these measures; however, these rules of thumb can be misleading [122]. Instead, we recommend that instructors and researchers use the 75% threshold to understand how courses either do or do not support students in developing the attitudes necessary to become physicists and if those courses repay, maintain, or exacerbate the educational debts society owes to students from marginalized groups.

## VIII. CONCLUSION

The American Physical Society (APS) statement on diversity [2] takes a strong stance on creating educational and professional systems that support the success of students from diverse background. In it the APS, "call[s] upon policymakers, administrators, and managers at all levels to enact policies and promote budgets that will foster greater diversity in physics. … We call upon the physics community as a whole to work collectively to bring greater diversity wherever physicists are educated or employed." If few women and BIPOC students start college physics courses with the attitudes the physics community requires to become a professional physicist and physics instruction continues to not support students in developing these attitudes, physics will not become more diverse. Pedagogies with an epistemilogical or model-based focus can support students in developing the attitudes necessary for becoming a physicist [33]. In particular, Traxler and Brewe [14] found Modeling Instruction supports women and BIPOC students in developing more expertlike attitudes. Meeting the call of the APS requires researchers to confirm these results in other settings and to further identify the mechanisms by which they support students changing their attitudes. Meeting this call will also require faculty and instructors demanding and administrators supplying the resources to implement evidence-based pedagogies to support diverse students in developing expertlike attitudes.

Physics is a cultural endeavor. What we study, how we study it, the language and symbols we use, and who gets to do the studies are dictated by cultural practices. Physicists sometimes describe physics as a culture of no culture [123] and contend physics is rooted in the natural world, is empirical, and is above culture [124]. But if physics is a culture of no culture, why is physics so homogenous? As Prescod-Weinstein argues [124], "Identity should not matter where there is truly no 'culture,' and anyone noticing the homogeneity of the community will then experience a cognitive conflict." Requiring this set of expertlike attitudes for students to become physicists and failing to provide instruction that develops those attitudes are cultural practices. Individual members within the physics community can take steps in their teaching, mentoring, and outreach to change these cultural practices and meet the call of the APS.

The physics community can support more students from marginalized groups becoming physicists by addressing the systemic and value laden barriers society and the physics culture creates for these students. A lack of support in developing expertlike attitudes is not the only barrier students face in becoming physicists. To identify and address these barriers, we need to examine our cultural practices within physics. Many of the attitudinal norms and cultural practices in physics, such as competition, individualism, and solitary practice, have greater costs for women and BIPOC students [29,39]. Focusing on attitude development without evaluating those attitudes is a form of assimilation rather than education. Assimilation blames people from marginalized groups for the injustices they bear and is racist and sexist [82]. However, to not address society's educational debts owed to students due to racism and sexism is also racist and sexist. Here in we risk being caught in a catch-22 that leads to inaction. Educators who adopt pedagogies that support attitude development can have a positive impact on who gets to pursue becoming a physicist. Increased diversity in physics may shift the status quo of what it means to be and to become a physicist. But we do not have to wait to change the status quo. We must make the effort now to understand the cultural practices of physics, how those practices support or hinder scientific progress, and how those practices enact or fight injustices. Our physics community can develop resources and tools to transform our cultural practices to be just. Changing the status quo of what it means to be and do physics is equally important and mutually supporting to changing the status quo of who gets to do physics by addressing the educational debts society owes students from marginalized groups [86].

In pursuing equity, critical race theory warns that interventions to support women and BIPOC students will only be implemented when they also benefit the group in power, White men, thereby creating an *interest* *convergence*. The physics community should attend to which institutions and students receive resources to improve diversity in physics and are represented in investigations of those interventions. We must not only create, and test pedagogies that address society's educational debts across a wide range of settings, we must also investigate the extent to which those pedagogies are used in environments where they benefit marginalized students. Evidence-based pedagogies that are disproportionately used at primarily White, research-intensive, four-year universities enact systemic racism and classism.

## IX. LIMITATIONS

Our data do not represent the breadth of introductory mechanics courses and therefor limits the generalizability of our findings. Only two institutions in the data set were two-year colleges, but a large portion of college physics courses are taken at two-year colleges [101]. Future work should explore issues of attitudes and equity across the intersection of race and gender in the two-year college environment. Similarly, research should seek to replicate findings of pedagogies that support students in developing their attitudes at two-year colleges.

The CLASS was developed using data from two institutions and based on interviews with three physicists. Its development did not account for the variety of attitudes about learning and doing physics held by a diverse sample of professional physicists. It is possible that the narrow sample used in the development of the CLASS causes it to miss attitudes that students from marginalized groups hold that are valued by the physics community. International researchers have, however, translated the CLASS into eight languages other than English. These translations indicate that the attitudes measured by the CLASS are valuable to a physicists from across the world. Future work investigating students physics attitudes would benefit from a deeper and wider understanding of the attitudes that physicists hold. Potential future research questions include (i) how do these attitudes vary across individual physicists and communities within physics? and (ii) what are the costs and benefits of these attitudes for individuals and the field as a whole?

## ACKNOWLEDGMENTS

We are grateful to the Learning Assistant Program at the University of Colorado Boulder for establishing the LASSO platform. This work is funded in part by NSF-IUSE Grant No. DUE-1525338 and NSF-HSI Grant No. DUE-1928596 and is Contribution No. LAA-065 of the Learning Assistant Alliance. The data used in the analysis for this study came from the LASSO Platform: LASSO data 6 19.

## APPENDIX

### A. Descriptive statistics

Table V breaks down the descriptive statistics across demographic groups used in the final models. Figure 4

TABLE V. Descriptive statistics by race, gender, and course type.

| | | Algebra-based | | | | | | Calculus-based | | | | | |
|---|---|---|---|---|---|---|---|---|---|---|---|---|---|
| | | | Pre | | Post | | Gain | | Pre | | Post | | Gain |
| Race | Gender | N | Mean | S.D. | Mean | S.D. | Mean | S.D. | N | Mean | S.D. | Mean | S.D. | Mean | S.D. |
| All | All | 2503 | 56.2 | 16.3 | 55.6 | 17.4 | −0.6 | 13.3 | 2170 | 63.7 | 15.2 | 60.9 | 15.2 | −2.8 | 10.4 |
| All | Women | 1462 | 54.9 | 16.2 | 53.8 | 17.1 | −1.1 | 13.3 | 697 | 61.9 | 15.4 | 58.8 | 15.2 | −3.1 | 10.5 |
| All | Other | 33 | 55.4 | 20.1 | 58.5 | 18.3 | 3.1 | 17.3 | 29 | 58.0 | 23.1 | 56.7 | 21.9 | −1.3 | 12.6 |
| All | Men | 1008 | 58.1 | 16.2 | 58.1 | 17.4 | 0.0 | 13.2 | 1444 | 64.7 | 14.8 | 62.0 | 15.0 | −2.7 | 10.2 |
| Asian | Women | 309 | 53.8 | 15.1 | 52.9 | 16.5 | −0.9 | 12.8 | 154 | 58.2 | 15.2 | 56.9 | 14.4 | −1.3 | 10.5 |
| Asian | Men | 166 | 57.9 | 17.1 | 54.5 | 17.2 | −3.4 | 12.5 | 270 | 61.7 | 15.0 | 59.9 | 15.3 | −1.8 | 10.4 |
| Asian Hispanic | Women | 5 | 53.9 | 12.7 | 44.0 | 10.3 | −9.9 | 4.5 | 4 | 47.9 | 15.8 | 48.5 | 15.6 | 0.6 | 8.3 |
| Asian Hispanic | Men | 4 | 50.8 | 19.8 | 49.4 | 12.4 | −1.4 | 20.0 | 7 | 53.4 | 17.2 | 42.2 | 18.5 | −11.2 | 20.9 |
| Black | Women | 98 | 51.5 | 15.2 | 48.1 | 15.4 | −3.5 | 12.6 | 42 | 58.1 | 12.5 | 51.9 | 15.8 | −6.2 | 12.6 |
| Black | Men | 40 | 52.6 | 15.8 | 53.3 | 16.4 | 0.8 | 8.8 | 63 | 60.6 | 13.6 | 55.5 | 15.6 | −5.1 | 8.7 |
| Black Hispanic | Women | 8 | 42.1 | 12.2 | 43.8 | 15.1 | 1.7 | 9.0 | 7 | 53.6 | 16.1 | 46.8 | 12.6 | −6.7 | 10.4 |
| Black Hispanic | Men | 3 | 70.1 | 6.7 | 65.2 | 5.1 | −4.8 | 11.7 | 5 | 68.8 | 17.7 | 61.3 | 14.3 | −7.5 | 7.8 |
| Hispanic | Women | 65 | 54.2 | 14.2 | 56.8 | 14.8 | 2.7 | 11.1 | 43 | 56.0 | 16.4 | 53.4 | 13.0 | −2.6 | 11.2 |
| Hispanic | Men | 58 | 53.8 | 14.7 | 52.0 | 16.7 | −1.8 | 14.0 | 114 | 58.4 | 13.5 | 55.3 | 13.3 | −3.2 | 9.8 |
| Other | Women | 96 | 52.8 | 16.7 | 51.5 | 17.2 | −1.3 | 15.6 | 55 | 62.1 | 16.3 | 60.3 | 15.0 | −1.9 | 11.8 |
| Other | Men | 71 | 57.6 | 18.1 | 57.3 | 17.4 | −0.3 | 14.9 | 105 | 62.0 | 15.4 | 60.8 | 15.2 | −1.2 | 10.3 |
| White | Women | 760 | 56.4 | 16.9 | 55.2 | 18.0 | −1.2 | 13.6 | 344 | 65.1 | 14.9 | 61.4 | 15.2 | −3.7 | 10.2 |
| White | Men | 583 | 59.3 | 15.6 | 60.3 | 17.4 | 1.0 | 13.4 | 726 | 67.6 | 14.4 | 64.9 | 14.3 | −2.8 | 10.1 |
| White Hispanic | Women | 121 | 53.7 | 14.5 | 52.8 | 14.6 | −0.9 | 12.1 | 48 | 61.3 | 14.5 | 58.3 | 15.0 | −3.0 | 9.0 |
| White Hispanic | Men | 83 | 56.5 | 16.9 | 57.7 | 16.6 | 1.1 | 11.6 | 154 | 64.9 | 13.6 | 61.3 | 14.6 | −3.6 | 10.5 |

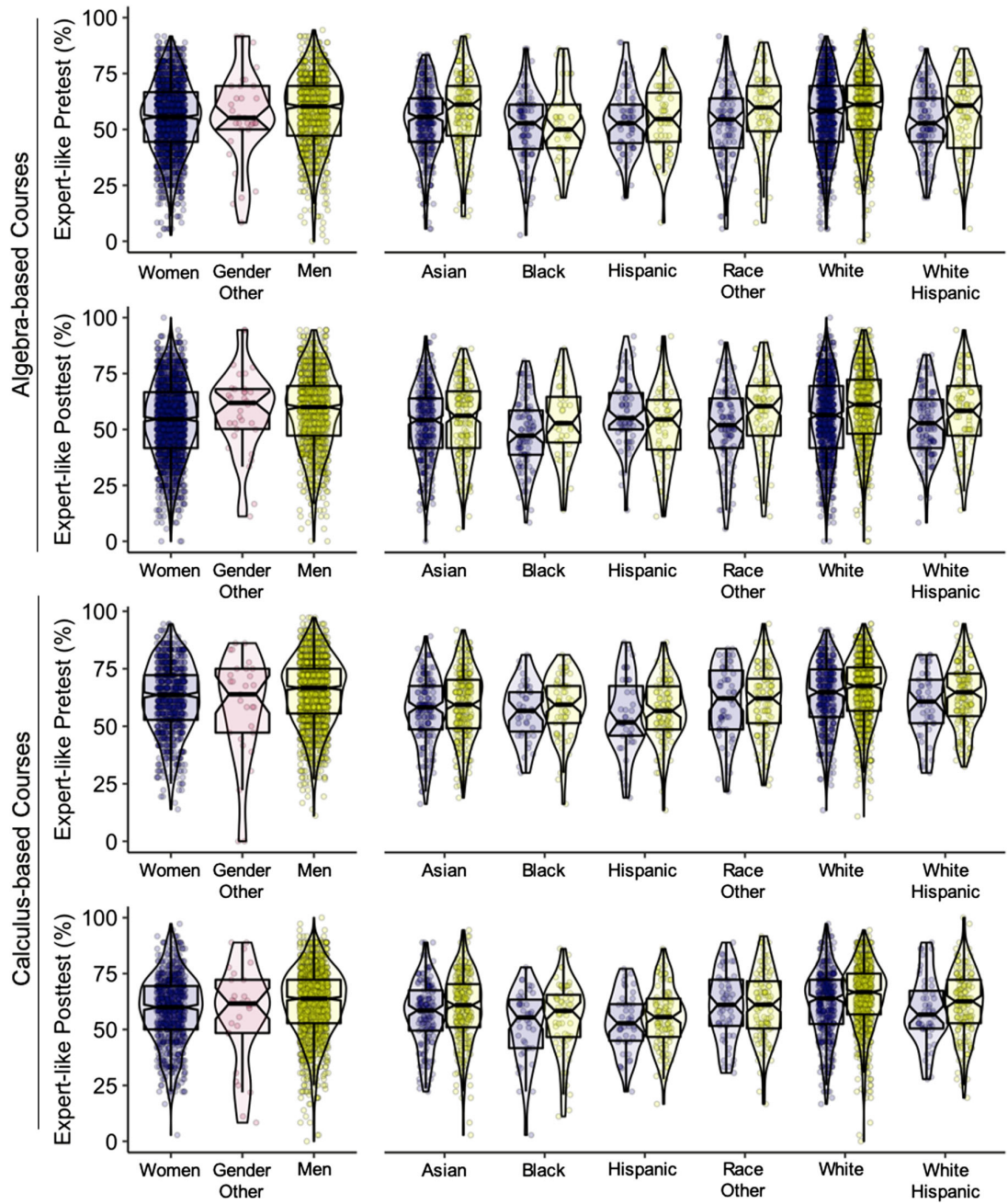

FIG. 4. Violin plots, box plots, and scatter plots of the data for pretest and post-test CLASS scores in algebra- and calculus-based courses. The violin plot is a reflected density plot to show the distribution of the data. The notched box plots show the distribution but focus attention on the medians with notches to show the 95% confidence intervals. The scatter plot is jittered to randomly distribute the points left and right of the vertical axis for clarity. The scatter plot illustrates the number of data points in each group and adds detail to how the data are distributed, particularly in the tails. These plots show easily identifiable differences across groups, but also show these differences are shifts in the distributions of scores and not gaps that separate groups. Many of the plots illustrate a small negative (downward) skew with more very low scores than would occur in a truly normal distribution.

includes violin plots, box plots, and jittered scatter plots. These figures used the average values for each student for all ten imputed datasets. We included these plots because they provide data transparency for readers and because they break down the "gap gazing" perspective that means and standard errors from either descriptive statistics or statistical models can reinforce. In other words, the data show differences in the mean scores between groups (intergroup variance) but the spread of scores within each group (intragroup variance) is much larger, which is shown by the overlap in the distributions across all groups. Across all the plots in Fig. 4, several features are worth noting: differences but not gaps across groups, medians that tend to be above 50%, and a negative skew (downward) for many of the distributions.

TABLE VI. Hierarchical linear model coefficients and standard errors for predicted average attitudes.

| | Algebra-based | | | | Calculus-based | | | |
|---|---|---|---|---|---|---|---|---|
| | Pre | | Post | | Pre | | Post | |
| | β | SE | β | SE | β | SE | β | SE |
| Intercept | 59.6 | 4.7 | 60.6 | 5.2 | 61.9 | 4.0 | 62.6 | 4.4 |
| Gender other | −1.3 | 3.2 | 1.9 | 3.4 | −5.5 | 3.0 | −4.5 | 3.6 |
| Hispanic | −4.0 | 4.4 | −6.0 | 5.0 | −3.4 | 3.7 | −7.5 | 4.1 |
| White | −0.2 | 4.7 | −0.8 | 5.2 | 5.1 | 3.9 | 2.0 | 4.5 |
| Women | −2.9 | 2.2 | −1.8 | 2.6 | −0.9 | 2.2 | −1.4 | 2.9 |
| Black | −4.8 | 5.2 | −5.1 | 5.5 | −0.5 | 4.1 | −6.1 | 5.0 |
| Asian | −2.2 | 4.7 | −5.8 | 5.3 | −1.6 | 4.0 | −3.5 | 4.6 |
| Race other | −3.3 | 4.6 | −5.6 | 5.1 | 0.0 | 4.0 | −1.6 | 4.7 |
| Hisp.*White | 0.3 | 4.4 | 2.1 | 4.9 | 0.9 | 3.7 | 4.1 | 4.4 |
| Women*Black | −0.6 | 3.7 | −5.0 | 4.4 | −2.7 | 3.8 | −2.9 | 4.7 |
| Women*Asian | −0.9 | 2.8 | 0.1 | 3.1 | −3.2 | 2.7 | −1.6 | 3.4 |
| Women*Hispanic | 1.3 | 2.1 | 2.8 | 2.4 | −1.4 | 2.1 | 0.4 | 2.7 |
| Women*White | −0.2 | 2.2 | −3.1 | 2.7 | −2.2 | 2.3 | −2.3 | 2.9 |

### B. Model outputs

Tables VI and VII present the model outputs for all models presented in the article. Table VI presents the model coefficients and standard errors used to generate the predicted average attitudes. Table VII presents the model coefficients and standard errors in logits, which we converted to probabilities for the proportions of students above the 75% cutoff.

### C. Assumption checking

We are unaware of any single uniformly agreed to method for pooling the test results of the assumption checking for multilevel models when researchers use multiple imputation [108]. We performed the assumption checks using each imputed dataset. We present the results for the assumption checking using the pooled dataset made by averaging all of the imputed datasets. The pooled dataset on its own should not be used for checking the assumptions. We are, however, using it because our conclusions across all of the imputed models aligns with the results from the pooled data and to greatly simplify presentation of the assumption checking. To test the assumption of linearity, we plotted the residual variance against the fitted values, shown in Figs. 5 and 6. In our visual inspection of the figures we saw no obvious trends and concluded that the model met the assumption of linearity. To test for homogeneity of variance we created a box plot of the residuals across courses, shown in Figs. 5 and 6, and performed an ANOVA of the residuals across courses. A visual inspection of the box plot showed the courses' residuals had consistent medians and interquartile ranges and therefore met the assumption of homogeneity of variance. The ANOVA supported our visual check because it did not find a statistically significant difference ($p > 0.05$) in the variances across courses. Finally, we visually checked the assumption of normality of residuals using a qq plot of the observed and expected values, shown in Figs. 5 and 6. The small negative curvature in the qq plots indicated a small leftward skew in the residuals indicating there are more large negative residuals than a normal distribution would produce. This likely occurred because the data, as shown in the violin plots Fig. 4 tends to have a slight left (down) skew. Hence the model is overdriven by lower test scores. Gelman and Hill [125] point out that meeting the assumption of normally distributed residuals is of little importance to the regression line. The small skew in the residuals could have a very small effect on the standard errors. We expect that this skew had no effect on our conclusions for two reasons. First, the residuals were not meaningfully correlated with any demographic variable, $r \sim 0$. Second, we did not use $p$ values as binary indicators of significance; therefore, any small shifts in the standard errors would have minimally changed our conclusions.

TABLE VII. Hierarchical generalized linear model outputs for the pretest 75% cutoff models.

| Race | Gender | Algebra-based | | Calculus-based | |
|---|---|---|---|---|---|
| | | Est. | SE | Est. | SE |
| Asian | Women | −2.13 | 0.20 | −1.86 | 0.25 |
| | Men | −1.50 | 0.21 | −1.31 | 0.17 |
| Black | Women | −2.61 | 0.43 | −2.27 | 0.55 |
| | Men | −1.41 | 0.40 | −1.27 | 0.33 |
| Hispanic | Women | −2.50 | 0.41 | −1.80 | 0.37 |
| | Men | −2.18 | 0.41 | −1.93 | 0.28 |
| White | Women | −1.58 | 0.11 | −0.92 | 0.15 |
| | Men | −1.38 | 0.11 | −0.56 | 0.11 |
| White Hispanic | Women | −2.26 | 0.30 | −1.43 | 0.32 |
| | Men | −1.67 | 0.29 | −0.98 | 0.20 |

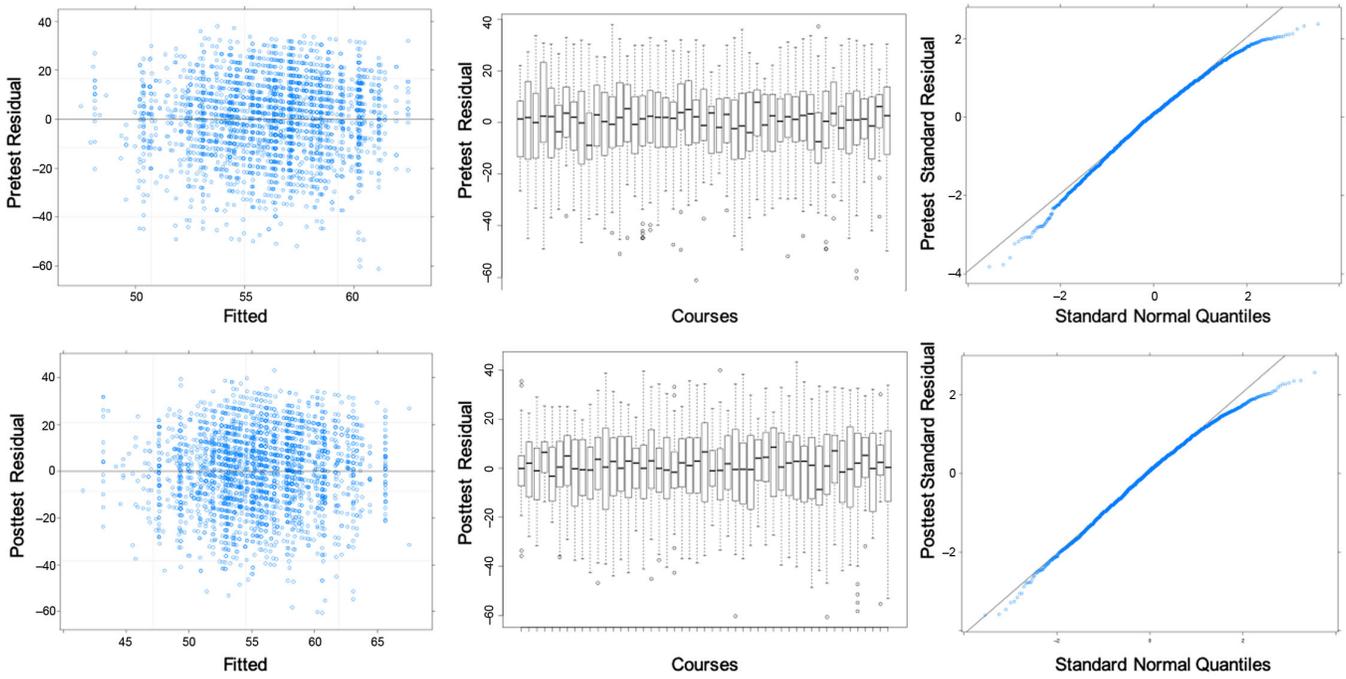

FIG. 5. Assumption checking for the pretest and post-test models for algebra-based physics courses.

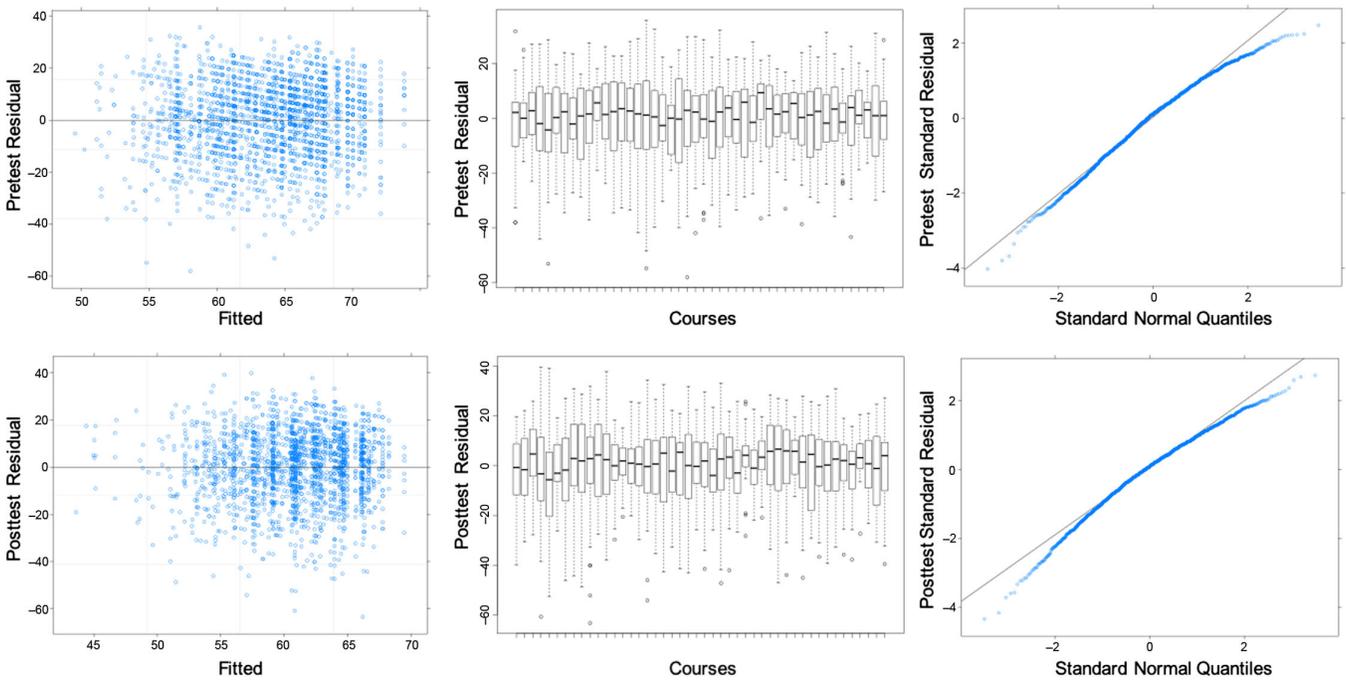

FIG. 6. Assumption checking for the pretest and post-test models for calculus-based physics courses.